\documentclass{article}

\usepackage{graphicx}
\usepackage{subfigure}
\usepackage{xcolor}
\usepackage{geometry}
\usepackage{caption}
\usepackage{amsmath}
\usepackage{comment}
\usepackage[colorlinks=true,linkcolor=black,citecolor=blue]{hyperref}
\usepackage{algorithm, algpseudocode, algorithmicx}

\newtheorem{thm}{Theorem}[section]
\newtheorem{defn}[thm]{Definition}

\usepackage{multirow}
\title{Strongly Chordal Graph Generation using Intersection Graph Characterization}
\author{Md Zamilur Rahman \\
Asish Mukhopadhyay \\
School of Computer Science \\
University of Windsor\\
Canada}
\date{}

\begin{document}
\maketitle{}

\section{Introduction}\label{intro}
 
Strongly chordal graphs, introduced by Farber~\cite{DBLP:journals/dm/Farber83}, are a subclass 
of chordal graphs. He also established a number of different characterizations for this class of 
graphs. These include an intersection graph characterization~\cite{FarberThesis} that is 
analogous to a similar characterization for chordal graphs~\cite{GAVRIL197447}. Seker 
et al. \cite{DBLP:journals/corr/abs-1810-13326} exploited this characterization of chordal graphs to obtain an algorithm for generating them. In this paper, we propose an algorithm to show that strongly chordal graphs
can also be generated using their intersection graph characterization.\\

The following essential definitions from Farber \cite{DBLP:journals/dm/Farber83} underlie this 
characterization. Let $r$ be the root $T$ of an edge-weighted tree. The edge-weights are 
positive numbers that can be conveniently interpreted as the lengths of the edges. 

\begin{defn}{\label{def1}}
The weighted distance from a node $u$ to a node $v$ in $T$,  denoted by $d_T^*(u,v)$, is the sum of the lengths of the edges of the (unique) path from $u$ to $v$.
\end{defn}

\begin{defn}{\label{def2}}
Let $T_1$ and $T_2$ be two subtrees of $T$. Subtree $T_1$ is full with respect to $T_2$, denoted by $T_1>T_2$, if for any two vertices $u,v \in T_2$ such that $d_T^*(r,u)\leq d_T^*(r,v)$, $v\in V(T_1)$ implies that $u\in V(T_1)$.
\end{defn} 

\begin{defn}{\label{def3}}
A collection of subtrees $\{T_0, T_2,\dots, T_{k - 1}\}$ of $T$ is compatible if for each pair of 
subtrees $T_i$ and $T_j$ either $T_i>T_j$ or $T_j>T_i$. 
\end{defn}

Using the definitions above, Farber established the following intersection graph characterization for strongly chordal graphs. 

\begin{thm}{\rm~\cite{FarberThesis}}{\label{theTheorem}}
	A graph is strongly chordal if and only if it is the intersection graph of a compatible collection of subtrees of a rooted, weighted tree, $T$.
\end{thm}

\section{The Algorithm}

Let $S$ be an adjacency matrix whose rows correspond to a compatible collection
of subtrees, $\{T_0, T_2, \ldots, T_{k - 1}\}$, of a rooted, weighted tree $T$ as in 
Theorem~\ref{theTheorem} and columns correspond to the vertices $\{v_0, v_2, \ldots, v_{n - 1}\}$ 
of $T$, arranged from left to right in order of non-decreasing distance from the root, $r$. 
Our main observation is that Definition~\ref{def2} can be re-interpreted to imply that 
the matrix $S$ cannot have 
$\Delta_1 = \bigl[ \begin{smallmatrix}1 & 1\\ 0 & 1\end{smallmatrix}\bigr]$ as a sub-matrix. 
More precisely, if $i$ and $j$ are two rows of $S$, corresponding 
to compatible subtrees $T_i$ and $T_j$ of $T$, then there cannot exist columns $k$ and $l$ 
that intersect these two rows to create $\Delta_1$. 
Thus $S$ belongs to the class of 
0-1 matrices that do not have $\Delta_1$ as a submatrix. Note that this is only a necessary condition. If we can generate a 0-1 matrix that satisfies this necessary condition, we have to 
further ensure that each row corresponds to a subtree of a weighted tree $T$. The details of how 
this can be achieved are described in the algorithm below that is built atop our observation
of the forbidden sub-matrix property of $S$.\\

Each of the entries of the first row and the first column are randomly set to $0$ or $1$. 
The entries of the submatrix [$1..k - 1, 1..n - 1$] 
 are carefully set to $0$ or $1$ so as not have $\Delta_1$ as a submatrix. 
This is done in row major order. While setting the entry of the $i$-th row and $j$-th column
we check exhaustively the entries in the columns to the left of the $j$-th column and the 
entries above the $i$-th row to make sure that no $2 \times 2$ submatrix is equal to $\Delta_1$.
To have a compatible collection of subtrees $\{T_0, T_2,\dots, T_{k-1}\}$ of a given tree $T$, we also do not want 
$\Delta_2 = \bigl[ \begin{smallmatrix}0 & 1\\ 1 & 1\end{smallmatrix}\bigr]$ as a submatrix. 
Such a submatrix can create cycles in the tree $T$ we wish to construct from the rows 
of our matrix representing a collection of compatible subtrees. Algorithm~\ref{algosmat} generates a $0-1$ matrix $S$ without $\Delta_1$ or $\Delta_2$ as a submatrix.
  
\begin{algorithm}[htb]
	\caption{$0-1$ Matrix Generation}\label{algosmat}
	\begin{algorithmic}[1]
		\Require The number of columns (nodes) $n$ and the number rows (subtrees) $k$ 
		\Ensure A matrix $S$
		\For{$i \gets 0$ to $k - 1$}
			\For{$j \gets 0$ to $n - 1$}
			\State $randomEntry \leftarrow random(0, 1)$ \Comment randomly chosen either $0$ or $1$
				\If {$i == 0$ or $j == 0$}
					\State $S[i][j] \leftarrow randomEntry$
				\Else
					\If {$randomEntry == 1$}
						\If {$(S[i - 1][j - 1] == 1$ and $S[i - 1][j] == 1$ and $S[i][j - 1] == 0)$ \textbf{or} $(S[i - 1][j - 1] == 0$ and $S[i - 1][j] == 1$ and $S[i][j - 1] == 1)$}
							\State $S[i][j] \leftarrow 0$ \Comment switch 1 to 0
						\Else
							\State $S[i][j] \leftarrow randomEntry$ \Comment keep 1 as a valid entry
						\EndIf
					\Else
						\State $S[i][j] \leftarrow randomEntry$ \Comment keep 0 as a valid entry
					\EndIf
				\EndIf
			\EndFor
		\EndFor
	\end{algorithmic}
\end{algorithm}
In the next phase, we prune some of the rows of $S$.
First, we remove rows with all 0's. Then we remove duplicate rows (if any) because they 
produce identical subtrees and denote the reduced matrix by $S'$. In the next step, we generate a strongly chordal graph from the matrix $S'$. Each subtree (row) 
represents a vertex in the strongly chordal graph, and there is an edge between two vertices in 
the strongly chordal graph if $T_i\cap T_j\neq\emptyset$. 
Algorithm~\ref{algosubtrees} takes the number of columns (nodes) $n$ and the 
number of rows (subtrees) $k$ ($k\leq n$) as inputs and outputs a 
strongly chordal graph $G$. 

\begin{algorithm}[htb]
	\caption{Strongly chordal Graph Generation from Subtrees}\label{algosubtrees}
	\begin{algorithmic}[1]
		\Require The number of columns (nodes) $n$ and the number of rows (subtrees) $k$
		\Ensure A strongly chordal graph $G$
		\State Call $0-1$ matrix generation algorithm to generate a matrix without the sub matrix $\bigl[ \begin{smallmatrix}1 & 1\\ 0 & 1\end{smallmatrix}\bigr]$ and $\bigl[ \begin{smallmatrix}0 & 1\\ 1 & 1\end{smallmatrix}\bigr]$
		\State Remove any rows with zeros only
		\State Remove any duplicate rows
		\State Generate a strongly chordal graph from rows by computing the intersection of each row
	\end{algorithmic}
\end{algorithm}

In the following paragraphs, we explain on an example the strongly chordal graph generation process step-by-step. \\

\textbf{Example:} Algorithm~\ref{algosmat} generates the following $0-1$ matrix $S$ with 
$k = 12$ and $n = 12$.

\[
S=\begin{pmatrix}
1 & 0 & 0 & 0 & 1 & 0 & 0 & 0 & 1 & 0 & 0 & 1 \\
1 & 0 & 1 & 0 & 0 & 0 & 0 & 1 & 0 & 1 & 1 & 0 \\
1 & 1 & 0 & 1 & 0 & 1 & 1 & 0 & 0 & 0 & 0 & 0 \\
1 & 0 & 1 & 0 & 0 & 0 & 0 & 0 & 0 & 0 & 0 & 0 \\
0 & 0 & 0 & 0 & 0 & 0 & 0 & 0 & 0 & 0 & 0 & 0 \\
1 & 0 & 0 & 0 & 0 & 0 & 0 & 0 & 0 & 0 & 0 & 0 \\
0 & 0 & 0 & 0 & 0 & 0 & 0 & 0 & 0 & 0 & 0 & 0 \\
1 & 1 & 0 & 0 & 0 & 0 & 0 & 0 & 0 & 0 & 0 & 0 \\
0 & 0 & 0 & 0 & 0 & 0 & 0 & 0 & 0 & 0 & 0 & 0 \\
0 & 0 & 0 & 0 & 0 & 0 & 0 & 0 & 0 & 0 & 0 & 0 \\
0 & 0 & 0 & 0 & 0 & 0 & 0 & 0 & 0 & 0 & 0 & 0 \\
0 & 0 & 0 & 0 & 0 & 0 & 0 & 0 & 0 & 0 & 0 & 0
\end{pmatrix}\]

After removing 6 rows (row index: $4^{th}$, $6^{th}$, $8^{th}$, $9^{th}$, ${10}^{th}$, ${11}^{th}$) that have only zero entries, we get the following matrix $S^{\prime}$:

\[
S'=\begin{pmatrix}
1 & 0 & 0 & 0 & 1 & 0 & 0 & 0 & 1 & 0 & 0 & 1 \\
1 & 0 & 1 & 0 & 0 & 0 & 0 & 1 & 0 & 1 & 1 & 0 \\
1 & 1 & 0 & 1 & 0 & 1 & 1 & 0 & 0 & 0 & 0 & 0 \\
1 & 0 & 1 & 0 & 0 & 0 & 0 & 0 & 0 & 0 & 0 & 0 \\
1 & 0 & 0 & 0 & 0 & 0 & 0 & 0 & 0 & 0 & 0 & 0 \\
1 & 1 & 0 & 0 & 0 & 0 & 0 & 0 & 0 & 0 & 0 & 0 \\
\end{pmatrix}\]

\begin{figure}[htb]
	\centering
	\subfigure[{\em $T_0$} \label{Fig-SubTree-1}]{\includegraphics[scale=.65]{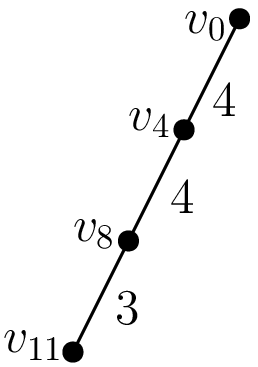}}\hspace{20pt}
	\subfigure[{\em $T_1$} \label{Fig-SubTree-2}]{\includegraphics[scale=.65]{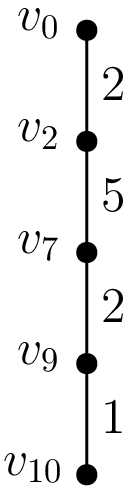}}\hspace{20pt}
	\subfigure[{\em $T_2$} \label{Fig-SubTree-3}]{\includegraphics[scale=.65]{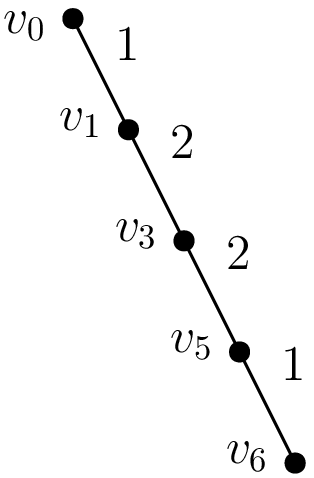}}\hspace{20pt}
	\subfigure[{\em $T_3$} \label{Fig-SubTree-4}]{\includegraphics[scale=.71]{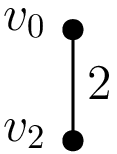}}\hspace{20pt}
	\subfigure[{\em $T_4$} \label{Fig-SubTree-5}]{\includegraphics[scale=.71]{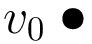}}\hspace{20pt}
	\subfigure[{\em $T_5$} \label{Fig-SubTree-6}]{\includegraphics[scale=.65]{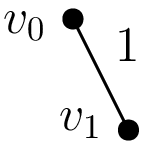}}\hspace{20pt}
	\caption{{\em Subtrees $\{T_0, T_1, T_2, T_3, T_4, T_5\}$}}
	\label{Fig-Subtrees}
\end{figure}

\begin{figure}[htb]
	\centering
	\includegraphics[scale = .65]{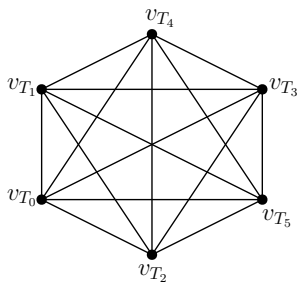}
	\caption{{\em A strongly chordal graph generated from subtrees $\{T_0, T_1, T_2, T_3, T_4, T_5\}$}}
	\label{Fig-SCG}
\end{figure}

From the matrix $S'$, we can see there are six subtrees. The subtrees are shown in Figure~\ref{Fig-Subtrees}. The strongly chordal graph shown in Figure~\ref{Fig-SCG} is generated by representing each $T_i$ of Figure~\ref{Fig-Subtrees} as a node $v_{T_i}$. There is an edge between $v_{T_i}$ and $v_{T_j}$ if the intersection of $T_i$ and $T_j$ is non-empty. \\

\section{Complexity}\label{Complexity}
Algorithm~\ref{algosmat} takes 
$O(n^4)$ time to generate matrix $S$, ensuring it does not have $\Delta_1$ or $\Delta_2$ as a submatrix. The intersection of two subtrees ($T_i\cap T_j$) can be computed in $O(n)$ time. Each subtree represents a vertex in a strongly chordal graph and if $T_i\cap T_j\neq\emptyset$, then there is an edge between two vertices in a strongly chordal graph. The insertion of an edge can be done in $O(1)$ time.

\section{Conclusions}\label{conclusions}
To the best of our knowledge this is the first algorithm for generating strongly chordal graphs based on an intersection graph characterization of this class. It would be interesting to improve on the time-complexity of this algorithm or find a more efficient way of generating strongly chordal graphs. We implemented this proposed algorithm in Python. As a matter of curiosity, we tested a large number of intersection graphs generated from $\Delta_1$-free matrices. Without exception, all of these passed the recognition algorithm test for strong chordality. It would be worthwhile to investigate this further. 

\end{document}